\documentclass[ALICE,manyauthors]{cernphprep}
\usepackage[comma,square,numbers,sort&compress]{natbib}
\usepackage{lineno}
\usepackage{xspace}
\usepackage{amsmath,amssymb}
\usepackage[comma,square,numbers,sort&compress]{natbib}

\usepackage{hyperref}

\usepackage{lineno}
\usepackage{esvect}
\usepackage{color}

\usepackage[T1]{fontenc}

\begin{document}
%

\newcommand{\pp}           {pp\xspace}
\newcommand{\ppbar}        {\mbox{$\mathrm {p\overline{p}}$}\xspace}
\newcommand{\XeXe}         {\mbox{Xe--Xe}\xspace}
\newcommand{\PbPb}         {\mbox{Pb--Pb}\xspace}
\newcommand{\pA}           {\mbox{pA}\xspace}
\newcommand{\pPb}          {\mbox{p--Pb}\xspace}
\newcommand{\AuAu}         {\mbox{Au--Au}\xspace}
\newcommand{\dAu}          {\mbox{d--Au}\xspace}

\newcommand{\s}            {\ensuremath{\sqrt{s}}\xspace}
\newcommand{\snn}          {\ensuremath{\sqrt{s_{\mathrm{NN}}}}\xspace}
\newcommand{\pt}           {\ensuremath{p_{\rm T}}\xspace}
\newcommand{\meanpt}       {$\langle p_{\mathrm{T}}\rangle$\xspace}
\newcommand{\ycms}         {\ensuremath{y_{\rm CMS}}\xspace}
\newcommand{\ylab}         {\ensuremath{y_{\rm lab}}\xspace}
\newcommand{\etarange}[1]  {\mbox{$\left | \eta \right |~<~#1$}}
\newcommand{\yrange}[1]    {\mbox{$\left | y \right |~<~#1$}}
\newcommand{\dndy}         {\ensuremath{\mathrm{d}N_\mathrm{ch}/\mathrm{d}y}\xspace}
\newcommand{\dndeta}       {\ensuremath{\mathrm{d}N_\mathrm{ch}/\mathrm{d}\eta}\xspace}
\newcommand{\avdndeta}     {\ensuremath{\langle\dndeta\rangle}\xspace}
\newcommand{\dNdy}         {\ensuremath{\mathrm{d}N_\mathrm{ch}/\mathrm{d}y}\xspace}
\newcommand{\Npart}        {\ensuremath{N_\mathrm{part}}\xspace}
\newcommand{\Ncoll}        {\ensuremath{N_\mathrm{coll}}\xspace}
\newcommand{\dEdx}         {\ensuremath{\textrm{d}E/\textrm{d}x}\xspace}
\newcommand{\RpPb}         {\ensuremath{R_{\rm pPb}}\xspace}
\newcommand{\RAA}         {\ensuremath{R_{\rm AA}}\xspace}

\newcommand{\nineH}        {$\sqrt{s}~=~0.9$~Te\kern-.1emV\xspace}
\newcommand{\seven}        {$\sqrt{s}~=~7$~Te\kern-.1emV\xspace}
\newcommand{\twoH}         {$\sqrt{s}~=~0.2$~Te\kern-.1emV\xspace}
\newcommand{\twosevensix}  {$\sqrt{s}~=~2.76$~Te\kern-.1emV\xspace}
\newcommand{\five}         {$\sqrt{s}~=~5.02$~Te\kern-.1emV\xspace}
\newcommand{\twosevensixnn}{$\sqrt{s_{\mathrm{NN}}}~=~2.76$~Te\kern-.1emV\xspace}
\newcommand{\fivenn}       {$\sqrt{s_{\mathrm{NN}}}~=~5.02$~Te\kern-.1emV\xspace}
\newcommand{\LT}           {L{\'e}vy-Tsallis\xspace}
\newcommand{\GeVc}         {Ge\kern-.1emV/$c$\xspace}
\newcommand{\MeVc}         {Me\kern-.1emV/$c$\xspace}
\newcommand{\TeV}          {Te\kern-.1emV\xspace}
\newcommand{\GeV}          {Ge\kern-.1emV\xspace}
\newcommand{\MeV}          {Me\kern-.1emV\xspace}
\newcommand{\GeVmass}      {Ge\kern-.2emV/$c^2$\xspace}
\newcommand{\MeVmass}      {Me\kern-.2emV/$c^2$\xspace}
\newcommand{\lumi}         {\ensuremath{\mathcal{L}}\xspace}

\newcommand{\ITS}          {\rm{ITS}\xspace}
\newcommand{\TOF}          {\rm{TOF}\xspace}
\newcommand{\ZDC}          {\rm{ZDC}\xspace}
\newcommand{\ZDCs}         {\rm{ZDCs}\xspace}
\newcommand{\ZNA}          {\rm{ZNA}\xspace}
\newcommand{\ZNC}          {\rm{ZNC}\xspace}
\newcommand{\SPD}          {\rm{SPD}\xspace}
\newcommand{\SDD}          {\rm{SDD}\xspace}
\newcommand{\SSD}          {\rm{SSD}\xspace}
\newcommand{\TPC}          {\rm{TPC}\xspace}
\newcommand{\TRD}          {\rm{TRD}\xspace}
\newcommand{\VZERO}        {\rm{V0}\xspace}
\newcommand{\VZEROA}       {\rm{V0A}\xspace}
\newcommand{\VZEROC}       {\rm{V0C}\xspace}
\newcommand{\Vdecay} 	   {\ensuremath{V^{0}}\xspace}

\newcommand{\ee}           {\ensuremath{e^{+}e^{-}}} 
\newcommand{\pip}          {\ensuremath{\pi^{+}}\xspace}
\newcommand{\pim}          {\ensuremath{\pi^{-}}\xspace}
\newcommand{\kap}          {\ensuremath{\rm{K}^{+}}\xspace}
\newcommand{\kam}          {\ensuremath{\rm{K}^{-}}\xspace}
\newcommand{\pbar}         {\ensuremath{\rm\overline{p}}\xspace}
\newcommand{\kzero}        {\ensuremath{{\rm K}^{0}_{\rm{S}}}\xspace}
\newcommand{\lmb}          {\ensuremath{\Lambda}\xspace}
\newcommand{\almb}         {\ensuremath{\overline{\Lambda}}\xspace}
\newcommand{\Om}           {\ensuremath{\Omega^-}\xspace}
\newcommand{\Mo}           {\ensuremath{\overline{\Omega}^+}\xspace}
\newcommand{\X}            {\ensuremath{\Xi^-}\xspace}
\newcommand{\Ix}           {\ensuremath{\overline{\Xi}^+}\xspace}
\newcommand{\Xis}          {\ensuremath{\Xi^{\pm}}\xspace}
\newcommand{\Oms}          {\ensuremath{\Omega^{\pm}}\xspace}
\newcommand{\degree}       {\ensuremath{^{\rm o}}\xspace}
\newcommand{\Dmeson}       {\ensuremath{\rm D^{0}}\xspace}
\newcommand{\Dbar}       {\ensuremath{\rm\overline{D}{}^0}\xspace}
\newcommand{\DDbar}       {\ensuremath{\Dmeson~{\rm and} ~\Dbar}\xspace}

\begin{titlepage}
\PHyear{2020}       
\PHnumber{085}      
\PHdate{19 May}  
\title{Elliptic flow of electrons from beauty-hadron decays in Pb--Pb collisions at $\sqrt{s_{\rm NN}}$ = 5.02 TeV}
\ShortTitle{Elliptic flow of beauty decay electrons}   

\Collaboration{ALICE Collaboration\thanks{See Appendix~\ref{app:collab} for the list of collaboration members}}
\ShortAuthor{ALICE Collaboration} 

\begin{abstract}

The elliptic flow of electrons from beauty hadron decays at midrapidity ($|y|$ $<$ 0.8) is
measured in Pb--Pb collisions at $\sqrt{s_{\rm NN}}$~= 5.02 TeV with the ALICE detector at
the LHC. 
The azimuthal distribution of the particles produced in the collisions can be parameterized with a Fourier expansion, in which the second harmonic coefficient represents the elliptic flow, $v_{\rm 2}$.
The $v_{\rm 2}$ coefficient of electrons from beauty hadron decays is measured for the first time in the transverse momentum (\pt) range 1.3--6 GeV/$c$ in the centrality class 30--50\%.
The measurement of electrons from beauty-hadron decays exploits their larger mean proper decay length $c\tau~\approx$ 500 $\mu$m compared to that of charm hadrons and most of the other background sources.
The $v_{\rm 2}$ of electrons from beauty hadron decays at midrapidity is found to be positive with a significance of 3.75$\sigma$. The results provide insights into the degree of thermalization of beauty quarks in the medium. A model assuming full thermalization of beauty quarks is strongly disfavoured by the measurement at high \pt, but is in agreement with the results at low \pt. Transport models including substantial interactions of beauty quarks with an expanding strongly-interacting medium describe the measurement within uncertainties. 

\end{abstract}
\end{titlepage}

\setcounter{page}{2} 


The main goal of the ALICE experiment~\cite{1748-0221-3-08-S08002} is the study of strongly-interacting matter at the high energy density and temperature reached in ultra-relativistic heavy-ion collisions at the Large Hadron Collider (LHC). In these collisions, the formation of a deconfined state of quarks and gluons, the quark--gluon plasma (QGP), is predicted by quantum chromodynamic (QCD) calculations on the lattice~\cite{pQCD1,pQCD2,pQCD3,pQCD4,pQCD5}. Because of their large masses, heavy quarks (charm (c) and beauty (b)) are mainly produced in hard scattering processes at the initial stage of the collision, before the formation of the QGP. Subsequently, they interact with the QGP, losing energy via radiative~\cite{Radiativea,Radiativeb} and collisional scattering~\cite{Colla,Collb,Collc} processes. Heavy-flavor hadrons and their decay products are thus effective probes to study the properties of the medium created in heavy-ion collisions. 
In non-central collisions, interactions among the medium constituents translate the initial spatial anisotropy in the coordinate space of nucleons participating in the collision into a momentum space anisotropy of produced particles in the final state~\cite{Snellings:2011sz}.
The momentum anisotropies are characterized by the flow harmonic coefficients $v_{\rm n}$
from the Fourier expansion of the particle azimuthal distribution with respect to  the azimuthal angle of the symmetry plane for the $n$-th harmonic.. The dominant flow harmonic is the elliptic flow $v_{\rm 2}$~\cite{Voloshin:1994mz}. 
At low transverse momentum, $p_{\rm T}~<$ 3 GeV/$c$, the measurements of positive $v_{\rm 2}$ are considered a manifestation of the collective hydrodynamical expansion of the medium~\cite{Heinz:2013th,Niemi:2015qia,Bernhard:2016tnd,Dubla:2018czx}. 
At high $p_{\rm T}$ ($p_{\rm T}~>$ 3 GeV/$c$), $v_{\rm 2}$ measurements give insight into the path-length dependence of the in-medium parton energy loss ~\cite{Pathlengtha,Pathlengthb, Noronha-Hostler:2016eow}.

The measurements of D-meson and J/$\psi$ $v_2$ in heavy-ion collisions, performed at RHIC in Au--Au collisions at $\sqrt{s_{\rm NN}}$ = 200 GeV \cite{Adamczyk:2017xur} and at the LHC in Pb--Pb collisions at $\sqrt{s_{\rm NN}}$ = 2.76 TeV and 5.02 TeV \cite{Acharya:2019mom, Abelev:2013lca, Acharya:2018bxo, Abelev:2014ipa, Acharya:2017qps, Sirunyan:2017plt, Acharya:2017tgv}, suggest that the interaction of charm quarks with the medium is sufficiently strong to make them thermalize and thereby take part in the collective flow of the medium \cite{Beraudo:2017gxw, Nahrgang:2013xaa, Song:2015ykw, Ke:2018jem, Katz:2019fkc, He:2011qa, Zigic:2018ovr}.
Additional mechanisms, like coalescence of charm quarks with the lighter quarks produced in the medium, can contribute to the flow of heavy-flavor particles~\cite{Recombinationa}.
Models consistent with the flow measurements of charm quarks have charm quark thermalization times of the order of the system lifetime ($\approx$ 10 fm/$c$) \cite{Beraudo:2017gxw}. This indicates that low-\pt charm quarks may be fully thermalized in the QGP due to their interactions with the medium. 
A non-thermalized probe is required to assess the interaction with the medium more thoroughly, with the heavier beauty quarks being a natural candidate.
It has been predicted by transport models that beauty quarks may experience sufficient scattering in the medium to produce a positive $v_{\rm 2}$ values~\cite{Batsouli:2002qf, He:2014cla, He:2011qa}.

Measurements of the anisotropic flow of leptons from combined charm and beauty hadron decays also showed that heavy quarks undergo significant rescattering in the medium and thus participate in its expansion \cite{Adamczyk:2014yew,Adare:2006nq, Adam:2016ssk, Aaboud:2018bdg}. However, strong conclusions about the dynamics of the beauty quark alone can not be drawn from those measurements, and separation of the charm and beauty contribution is necessary. Some anisotropic flow measurements of open-beauty do exist. The measured $v_{\rm 2}$ coefficient of the non-prompt J/$\psi$,  which is dominated by B$\rightarrow$J/$\psi$ carried out by the CMS collaboration is consistent with zero within large experimental uncertainties for \pt $>$ 3 GeV/$c$~\cite{Khachatryan:2016ypw}.
Recent measurements of the $v_{\rm 2}$ coefficient for $\Upsilon$(1S) by ALICE \cite{Acharya:2019hlv}, for \pt $<$ 15 GeV/$c$, are consistent both with zero and with the small value predicted by transport models~\cite{Du:2017qkv, Bhaduri:2018iwr} within uncertainties. Studies based on the Blast-Wave model show that, due to the large $\Upsilon$(1S) mass, even with full thermalization a sizeable elliptic flow would only be expected at \pt $>$ 10 GeV/$c$ \cite{Reygers:2019aul}. Hence lighter beauty hadrons, and their decay particles, would provide important additional information for the study of the interaction of beauty quarks with the medium. 
Recent ATLAS measurement of $v_2$ of muons from heavy-flavor hadron decays, including the separation of charm and beauty quark contributions, in Pb--Pb collisions at $\sqrt{s_\mathrm{NN}}$ = 5.02 TeV for \pt $>$ 4 GeV/$c$ revealed smaller flow
coefficients for muons from beauty hadron decays compared to those from charm hadrons~\cite{Aad:2020grf}.

In this Letter, the measurement of the $v_{\rm 2}$ of electrons (and positrons) from beauty hadron decays at midrapidity ($|y|~<$~0.8) in Pb--Pb collisions at $\sqrt{s_\mathrm{NN}}$ = 5.02 TeV recorded in 2018 with the ALICE detector is reported. The measurement is performed for the first time in the $p_{\rm T}$ interval 1.3 $<$ $p_{\rm T}$ $<$ 6 GeV/$c$.
The measurement is based on 77$\times 10^{6}$ minimum bias Pb--Pb collisions with a primary vertex reconstructed within $\pm$10 cm from the detector center ~\cite{Abelev:2014ffa} in the 30--50\% centrality interval.
Two forward and backward scintillator arrays (V0A and V0C) are used to determine the collision centrality~\cite{vZero, centralitypaper}.

Electron candidate tracks, reconstructed with up to 159 measurement points in the Time Projection Chamber (TPC) and up to 6 in the Inner Tracking System (ITS), are required to fulfill standard track selection criteria as listed in~\cite{Acharya:2019mom, Adam:2016wyz}.
To minimize the contribution of electrons from photon conversions in the detector material of the ITS and the fraction of tracks with misassociated hits, tracks are required to have associated hits in both Silicon-Pixel-Detector (SPD) layers, which constitute the two innermost layers of the ITS.
This requirement removes most electrons from photon conversion produced outside the first SPD layer from the track sample. However, in the high-multiplicity environment of heavy-ion collisions, some of these can be mis-associated with hits in the SPD layers produced by other particles.
Electron identification is done using the TPC and the Time of Flight detector (TOF)~\cite{Acharya:2019mom, Adam:2016wyz}. 
Electrons are identified by requiring the measured time-of-flight up to the TOF radius of 3.8 m on average to be within 3$\sigma$ of the expected value for electrons and their specific energy loss d$E$/d$x$ in the TPC to be within -1$\sigma$ and +3$\sigma$ with respect to the expected d$E$/d$x$ of electrons.

Electrons passing the track and identification selection criteria originate, besides from beauty-hadron decays, from Dalitz and di-electron decays of prompt light neutral mesons and charmonium states, photon conversions in the detector material, semi-leptonic decays of prompt-charm hadrons and decay chains of hadrons carrying a strange (or anti-strange) quark.
Measurements of electrons from beauty-hadron decays exploit their larger average impact parameter ($d_{0}$), defined as distance of closest approach to the primary vertex in the plane transverse to the beam line, compared to that of charm hadrons and most other background sources.
The sign of the impact parameter value is determined from the relative position of the track to the primary vertex, i.e. if the primary vertex is on the left- or right-hand side of the track with respect to the particle momentum direction in the transverse plane.
The impact parameter is multiplied with the sign of the particle charge and the magnetic field configuration~\cite{Adam:2016wyz}.
Electrons from photon conversions in the detector material are created at some distance from the primary vertex and in the direction of the photon. Their tracks bend away from the primary vertex, leading to an asymmetry with a mean impact parameter $d_{0}$ $<$ 0. This asymmetric impact parameter distribution allows for a better separation from the other electron sources, which are mostly symmetric around 0.

The momentum anisotropies are characterized by the flow harmonic coefficients $v_{\rm n}$ and azimuthal angle of the symmetry plane of $n$-th harmonic $\Psi_{\rm n}$ from the Fourier expansion of particle azimuthal distribution in the plane transverse to the beam direction~\cite{Acharya:2018zuq}.
The experimental estimate of the symmetry plane for the 2-nd harmonic of the collision-geometry in the azimuthal direction, the event plane $\Psi_{2}$, is determined using the signals produced by charged particles in the eight azimuthal sectors of
each V0 array. Non-uniformities in the V0 acceptance and efficiency are corrected for using the procedure described in~\cite{Selyuzhenkov:2007zi}.
 
The  $v_{\rm 2}\{\mathrm{EP}\}$ is  given by

\begin{equation}   
v_{\rm 2}\{\mathrm{EP}\} = \frac{1}{R_2} \frac{\pi}{4} \frac{N_{\rm in} - N_{\rm out}}{N_{\rm in} + N_{\rm out}},
\label{v2eqNinNout}
\end{equation}

where $N_{\rm in}$ and $N_{\rm out}$ are the number of beauty-decay electrons in two 90$^{\circ}$-wide intervals of $\Delta \varphi$ = $\varphi - \Psi_{2}$: in-plane ($-\frac{\pi}{4}$ $<$ $\Delta \varphi$ $<$ $\frac{\pi}{4}$ and $\frac{3\pi}{4}$ $<$ $\Delta \varphi$ $<$ $\frac{5\pi}{4} $) and out-of-plane ($\frac{\pi}{4}$ $<$ $\Delta \varphi$ $<$ $\frac{3\pi}{4}$ and $\frac{5\pi}{4}$ $<$ $\Delta \varphi$ $<$ $\frac{7\pi}{4} $), respectively. The resolution ($R_{2}$) of the event plane is measured with the three sub-event method~\cite{Abelev:2014ipa}. The sub-events are defined according to the signals in the V0 detectors (both A and C sides) and the tracks in positive (0 $<$ $\eta$ $<$ 0.8)  and negative \mbox{($-$0.8 $<$ $\eta$ $<$ 0)} pseudorapidity regions of the TPC. $R_{2}$ is calculated in 1\% centrality intervals and a weighted average for the 30--50\% interval is obtained using the number of binary nucleon–nucleon collisions as weights~\cite{Abelev:2014ipa}. The average $R_{2}$ value in the 30--50\% centrality class is 0.77~\cite{Acharya:2018bxo}. 

The $N_{\rm in}$ and $N_{\rm out}$ yields of electrons from beauty-hadron decays are extracted by fitting the impact parameter distribution of all electron candidates in data with Monte Carlo (MC) templates for different electron sources~\cite{Adam:2016wyz}.
A MC sample of minimum-bias (MB) Pb--Pb collisions at $\sqrt{s_\mathrm{NN}}$ = 5.02 TeV, generated with HIJING v1.36~\cite{Hijing:ref}, is used to obtain the impact parameter distributions of photon conversions and Dalitz decays.
To increase the sample of electrons from charm- and beauty-hadron decays, a sample of charm and beauty quarks generated with PYTHIA6~\cite{Sjostrand:2006za} is embedded into each Hijing MC event.
The generated particles are propagated through the ALICE apparatus using GEANT3~\cite{Brun:1082634}.
Four classes of electron sources are used: electrons from beauty-hadron decays, from charm-hadron decays, from photon conversions, and from all other processes, the latter being dominated by Dalitz decays of light neutral mesons. 
The Dalitz decays of light neutral mesons happen essentially at the interaction vertex as does the production of most of the remaining hadron contamination. Thus, the impact parameter distributions of the reconstructed tracks are very similar and represent the \pt-dependent impact parameter resolution. The uncertainty due to the slight remaining difference is assessed by exchanging the Dalitz template for one of charged hadrons measured in data and scaling the resulting difference by an estimate of the contamination \cite{Acharya:2019mom} relative to the fraction of electrons from Dalitz decays. This results in an uncertainty of 0.009 on the final $v_2$ in the first \pt interval, falling quickly with \pt.
The small ($<< 1\%$) contributions from the decay of strange particles are accounted for by the Dalitz and conversion electron templates.
Due to the long lifetime, this contribution has a wide impact parameter distribution. 
The unaccounted contribution of
these strange decay electrons is negligible within the applied $d_{0}$ range of [-0.1, 0.1]
cm in the fitting procedure~\cite{Adam:2016wyz}.

\begin{figure}[ht!]
\centering
\includegraphics[width=0.8\columnwidth]{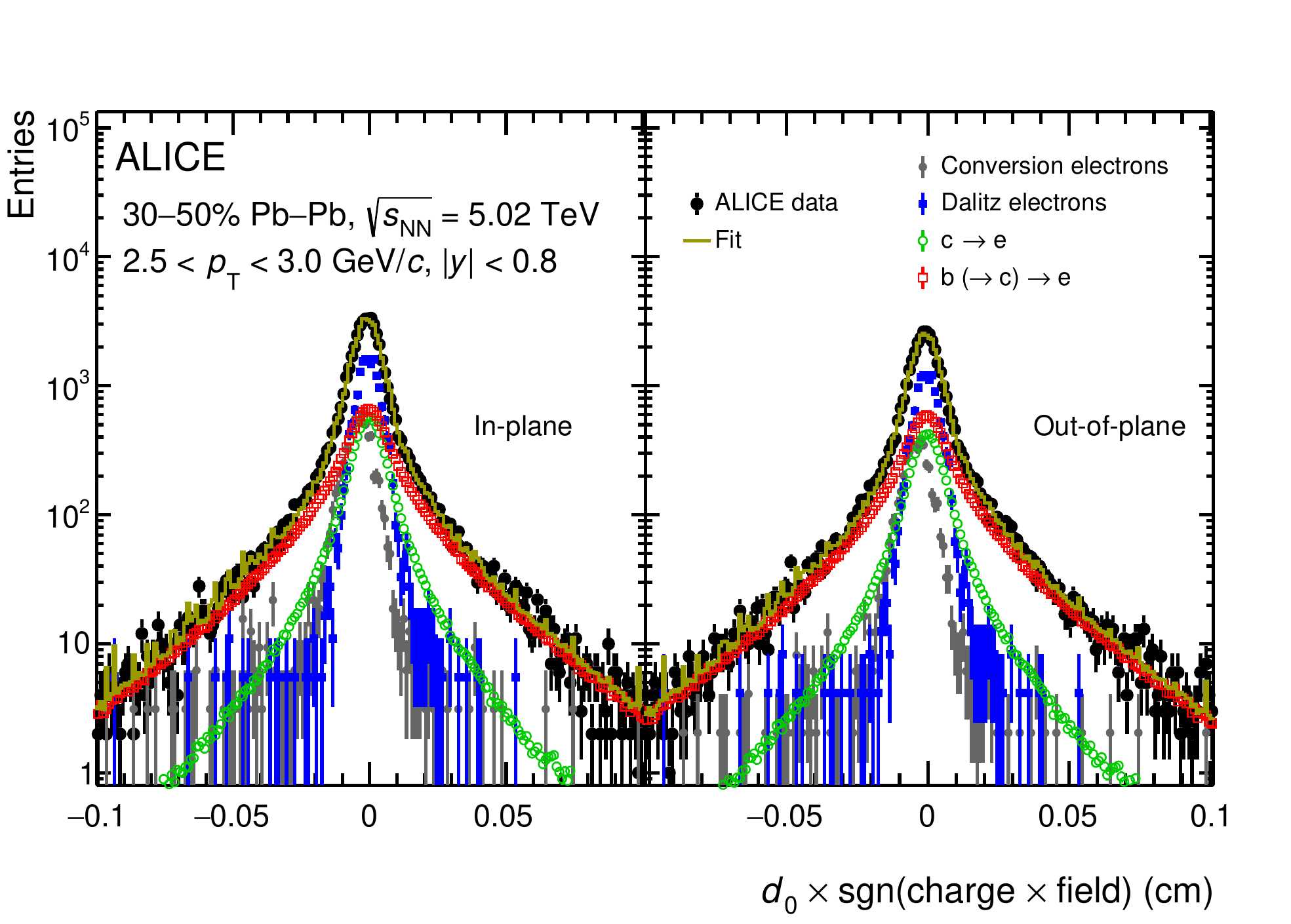}
\caption{Examples of the electron transverse impact parameter fits in-plane (left) and out-of-plane (right) for 2.5~$<$~\pt~$<$~3~GeV/$c$. Distributions from data and the four MC templates, electrons from beauty (b ($\rightarrow$ c) $\rightarrow$ e) and charm (c $\rightarrow$ e) hadron decays, electrons from photon conversions (Conversion electrons) and from other sources (Dalitz electrons) used in the fit are shown.
}
\label{doinplaneoutplane}
\end{figure}

The template fits are based on the method proposed in \cite{Barlow1993219} and implemented as in~\cite{Adam:2016wyz}. 
Figure \ref{doinplaneoutplane} shows examples of the resulting fits in in-plane (left panel) and out-of-plane (right panel) of electron $d_{0}$ distributions for the interval 2.5~$<$~\pt~$<$~3~GeV/$c$. In the figure the MC templates are corrected for all effects described in the following.

Detailed corrections to the MC templates are applied in order to take into account effects not simulated in MC. Special care is taken to assess differences in the in-plane and out-of-plane templates as the effects of the corrections do not cancel in the computation of the $v_{\rm 2}$.
The main corrections applied in the MC are: i) resolution of the $d_0$ distribution, ii) misassociated electrons from photon conversions and their multiplicity dependence, iii) \pt distribution of charm and beauty hadrons in in-plane and out-of-plane and iv) baryon-to-meson ratio of charm and beauty hadrons. A detailed description of these corrections are described below.

To ensure angular isotropy of the $d_0$ reconstruction in data, the mean $d_0$ of primary particles is compared in different regions in azimuth, $z$-position and \pt with a granularity smaller than the detector components and recentered prior to fitting.
Depending on \pt, the $d_0$ resolution in the MC simulations is about 11--13\% better
than in data~\cite{bib::Alice_performance, ALICE:2012ab} so this amount of smearing is added to make MC and data match. Primary pions and kaons are used for the comparison. It is observed that the resolution of the impact parameter does not depend significantly on the local track density.

The correct template shape of electrons from photon conversions depends on the production vertex and on the track multiplicity. In-plane and out-of-plane events have different local track densities, requiring separate corrections for the respective templates. 
This is achieved by choosing different centrality ranges for each template in the simulations. The ranges are defined based on how well they describe either the in-plane or out-of-plane reconstruction efficiencies of pions from $\rm K^0_S$ decays, as the production vertex of these decays is more accurately reconstructed. The systematic uncertainties are estimated by varying the nominal centrality classes in the simulations and are estimated to be 0.006 at low \pt and decreases to 0.001 with increasing \pt.

Because electrons of a given momentum from heavy-flavor hadron decays may originate from parent particles that have a broader momentum range, their $d_0$ distributions depend on the \pt distributions of these parent particles.
Hence it is necessary to correct for any difference in the \pt distribution of particles that decay to electrons between data and MC. For the charm case, this can be done by making use of the measured charm mesons \pt spectral shape and $v_2$ at the same collision energy~\cite{Acharya:2018hre,Acharya:2017qps}. From these measurements, separate \pt distributions and thus corrections are used for the in-plane and out-of-plane templates. To conservatively assess the uncertainty in the extracted $v_2$, the result is compared to a case where the assumed D meson $v_2$ is halved. An absolute systematic uncertainty of 0.004 is assigned from this comparison to the $v_2$ of electron from beauty hadron decay.

As there is no available measurement of the low-\pt beauty hadron $v_2$, the corrections for the beauty template are determined by using FONLL calculations~\cite{Cacciari:1998it} multiplied by \pt-dependent corrections that are estimated using a range of \RAA and $v_2$ values. 
The upper limit of the estimated \RAA value is the case of no suppression, \RAA~=~1, while the lower limit is obtained by interpolating the TAMU prediction~\cite{He:2014cla}, which is consistent with measurements of \RAA$\approx$~0.4 at high \pt~\cite{Adam:2016wyz}.
The arithmetic mean of the resulting $v_2$ values is used for the central points of the measurement, with the two limits used to estimate the systematic uncertainty. An absolute systematic variation of 0.0023 at low \pt and of 0.011 at high \pt is found and assigned as an uncertainty. 
A significant effect arises from the modification of the \pt spectra due to beauty-hadron $v_2$ since it gives a different correction for the in-plane and out-of-plane templates. For the central value of the measurement, the assumption of $v_2$ = 0.014$\times\pt^{2}$ $e^{(-1/3\times\pt)}$ (with \pt in units of GeV/$c$) is chosen as a generic function inspired by the prediction of the TAMU model \cite{He:2014cla}. The systematic uncertainty is conservatively evaluated by varying the $v_{\rm 2}$ value from zero to two times as large, the latter giving a peak of $0.14$. For these variations, the change in the measured beauty hadron decay electron $v_2$ is much smaller than the variation of the assumed hadron $v_2$. This gives a flat systematic uncertainty of 0.006 up to \pt = 4 GeV/$c$ and of 0.012 in the last \pt interval.

The impact parameter distribution of electrons from the different charm and beauty hadrons depends on the lifetime of these parent hadrons. Therefore, uncertainty in the relative contributions from different parents will translate into uncertainty in the $d_0$ distribution. For charm, the largest lifetime difference is in the decays of the baryons with respect to the mesons, while for beauty the lifetime of mesons and baryons are very similar and the effect of their different fractions in MC compared to data is negligible. 
For the charm case, a \pt-dependent correction is applied to the $\Lambda_{\rm c}$/$\rm D^{0}$ fraction based on model predictions~\cite{Andronic:2019wva, Plumari:2017ntm, Song:2018tpv}, which describe experimental measurements \cite{Acharya:2018ckj, Adam:2019hpq, Sirunyan:2019fnc}. 
This is compared to a \pt-independent correction that increases the $\Lambda_{\rm c}$/$\rm D^{0}$ by a factor of 3. The comparison shows no difference in $v_2$ for the two scenarios due to the effects cancelling out in the computation of the $v_2$.

The multiplicity dependence of the efficiency of the particle identification from the TOF detector is evaluated as in~\cite{Adam:2016wyz}. The efficiencies in-plane and out-of-plane are found to be within 0.5\% of each other, which results in an uncertainty of 0.0014 on the $v_2$.
No multiplicity dependence is found for the efficiency of particle identification from the TPC.

\begin{figure}[ht!]
\centering
\includegraphics[scale=0.5]{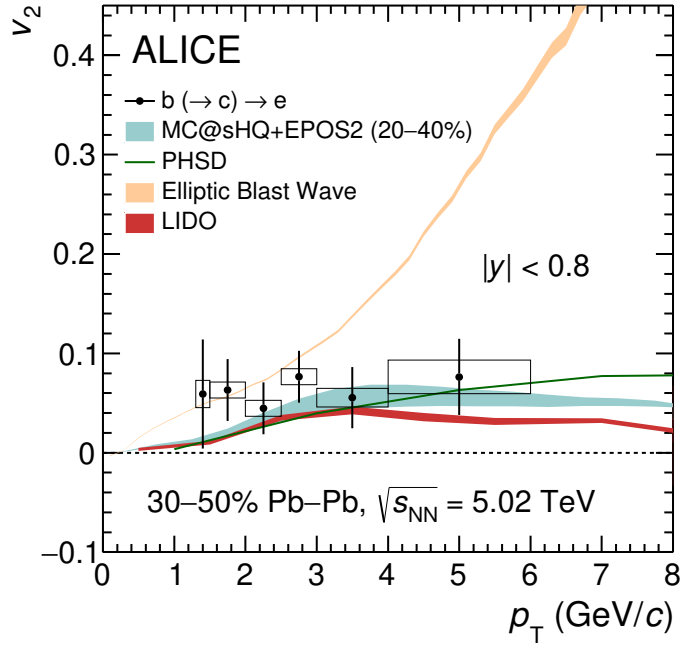}
\caption{Elliptic flow of electrons from beauty hadron decays in the 30--50\% centrality class in Pb--Pb collisions at $\sqrt{s_\mathrm{NN}}$ = 5.02 TeV at midrapidity as function of $p_{\rm T}$ compared with model calculations~ \cite{Nahrgang:2013xaa, Song:2015ykw,Ke:2018tsh, Ke:2018jem}.}
\label{btov2}
\end{figure}

Figure~\ref{btov2} shows the measured $v_{\rm 2}$ of electrons from beauty hadron decays at midrapidity ($|y|~<$ 0.8) as a function of $p_{\rm T}$ in Pb--Pb collisions at $\sqrt{s_\mathrm{NN}}$~ = 5.02 TeV in the 30--50\% centrality interval. A positive $v_{\rm 2}$ of electrons from beauty hadron decays with a significance of 3.75$\sigma$ is observed for the first time in this low \pt range (1.3--6 GeV/$c$) using the average $v_2$ divided by the uncertainty as a test statistic. The systematic uncertainties are assumed to be fully correlated for this purpose. No significant \pt dependence of the $v_2$ is observed.

The measured $v_{\rm 2}$ of beauty decay electrons is compared to the predictions from several transport models which include significant interaction of beauty quarks with a hydrodynamically-expanding QGP~\cite{Nahrgang:2013xaa, Song:2015ykw,Ke:2018tsh, Ke:2018jem}. 
These models are observed to well describe the D meson anisotropy and suppression in heavy-ion collisions at the LHC~\cite{Abelev:2013lca, Acharya:2018bxo, Abelev:2014ipa, Acharya:2017qps, ALICEDRPbPb, Adam:2015nna, Sirunyan:2017plt, Sirunyan:2017xss}.
The MC@sHQ+EPOS~\cite{Nahrgang:2013xaa} is a perturbative QCD model which includes radiative and collisional energy losses. The uncertainties of the model calculations are evaluated by varying from pure collisional to pure radiative energy losses, including also different scattering rates and different rescaling factors. Modification of nuclear parton distribution functions, due to effects from shadowing for example, is not considered for b quarks.
The LIDO model~\cite{Ke:2018tsh, Ke:2018jem} also includes both radiative and collisional energy loss.  This model uses experimental data to calibrate a Langevin-based transport model and extracts the transport coefficients directly from data via a Bayesian analysis.
In the case of LIDO, the reported model uncertainties are purely statistical. Within this model, the $v_2$ for beauty hadrons is much smaller than for charm hadrons.
The PHSD model~\cite{Song:2015ykw} is a microscopic off-shell transport model based on a Boltzmann approach which includes only collisional energy loss.
Initial-state event-by-event fluctuations are included in all transport models described here.
Even though the models differ in several aspects related to the interactions both in the QGP and in the hadronic phase as well as to the medium expansion, they all provide a fair description of the measurement. Similar agreement among these models was previously observed when compared to the \RAA of electrons from beauty-hadron decays~\cite{Adam:2016wyz}.
With the current experimental uncertainties, no model is clearly favoured or disfavoured.
A model calculation based on an extension of the blast-wave model~\cite{Reygers:2019aul} is also compared with the measurement. The calculation shown is based on $\rm B^0$ mesons, and the PYTHIA8 decayer is used for their decays into electrons~\cite{Sjostrand:2014zea}.
Assuming full thermalization, this model predicts a $v_2$ of $\Upsilon$(1S) close to zero in the range measured by ALICE, which is consistent with the measurement. The results for beauty hadron decay electrons give a much larger $v_2$ than that of $\Upsilon$(1S) due to the mass ordering effect. This shows that the measurement presented here is suitable to assess the degree of thermalization of beauty quarks at low \pt.
The error band represents purely the statistical uncertainty.
This simple model is qualitatively in agreement with the measurement within the uncertainties for \pt $<$ 3 GeV/$c$, while it significantly diverges from the data at higher \pt.

In summary, the measurement of the elliptic flow of electrons originating from beauty hadron decays at midrapidity in semicentral Pb--Pb collisions at $\sqrt{s_{\rm NN}}$ = 5.02 TeV is presented for the first time in this low \pt interval 1.3--6 GeV/$c$. The measurement is important for the understanding of the degree of thermalization of beauty quarks in the QGP. The $v_2$ of electrons from beauty hadron decays is found to be positive with a significance of 3.75$\sigma$. 
The measurement provides new insights and constraints to theoretical models of beauty quark interactions in the QGP.


\newenvironment{acknowledgement}{\relax}{\relax}
\begin{acknowledgement}
\section*{Acknowledgements}

The ALICE Collaboration would like to thank all its engineers and technicians for their invaluable contributions to the construction of the experiment and the CERN accelerator teams for the outstanding performance of the LHC complex.
The ALICE Collaboration gratefully acknowledges the resources and support provided by all Grid centres and the Worldwide LHC Computing Grid (WLCG) collaboration.
The ALICE Collaboration acknowledges the following funding agencies for their support in building and running the ALICE detector:
A. I. Alikhanyan National Science Laboratory (Yerevan Physics Institute) Foundation (ANSL), State Committee of Science and World Federation of Scientists (WFS), Armenia;
Austrian Academy of Sciences, Austrian Science Fund (FWF): [M 2467-N36] and Nationalstiftung f\"{u}r Forschung, Technologie und Entwicklung, Austria;
Ministry of Communications and High Technologies, National Nuclear Research Center, Azerbaijan;
Conselho Nacional de Desenvolvimento Cient\'{\i}fico e Tecnol\'{o}gico (CNPq), Financiadora de Estudos e Projetos (Finep), Funda\c{c}\~{a}o de Amparo \`{a} Pesquisa do Estado de S\~{a}o Paulo (FAPESP) and Universidade Federal do Rio Grande do Sul (UFRGS), Brazil;
Ministry of Education of China (MOEC) , Ministry of Science \& Technology of China (MSTC) and National Natural Science Foundation of China (NSFC), China;
Ministry of Science and Education and Croatian Science Foundation, Croatia;
Centro de Aplicaciones Tecnol\'{o}gicas y Desarrollo Nuclear (CEADEN), Cubaenerg\'{\i}a, Cuba;
Ministry of Education, Youth and Sports of the Czech Republic, Czech Republic;
The Danish Council for Independent Research | Natural Sciences, the VILLUM FONDEN and Danish National Research Foundation (DNRF), Denmark;
Helsinki Institute of Physics (HIP), Finland;
Commissariat \`{a} l'Energie Atomique (CEA) and Institut National de Physique Nucl\'{e}aire et de Physique des Particules (IN2P3) and Centre National de la Recherche Scientifique (CNRS), France;
Bundesministerium f\"{u}r Bildung und Forschung (BMBF) and GSI Helmholtzzentrum f\"{u}r Schwerionenforschung GmbH, Germany;
General Secretariat for Research and Technology, Ministry of Education, Research and Religions, Greece;
National Research, Development and Innovation Office, Hungary;
Department of Atomic Energy Government of India (DAE), Department of Science and Technology, Government of India (DST), University Grants Commission, Government of India (UGC) and Council of Scientific and Industrial Research (CSIR), India;
Indonesian Institute of Science, Indonesia;
Centro Fermi - Museo Storico della Fisica e Centro Studi e Ricerche Enrico Fermi and Istituto Nazionale di Fisica Nucleare (INFN), Italy;
Institute for Innovative Science and Technology , Nagasaki Institute of Applied Science (IIST), Japanese Ministry of Education, Culture, Sports, Science and Technology (MEXT) and Japan Society for the Promotion of Science (JSPS) KAKENHI, Japan;
Consejo Nacional de Ciencia (CONACYT) y Tecnolog\'{i}a, through Fondo de Cooperaci\'{o}n Internacional en Ciencia y Tecnolog\'{i}a (FONCICYT) and Direcci\'{o}n General de Asuntos del Personal Academico (DGAPA), Mexico;
Nederlandse Organisatie voor Wetenschappelijk Onderzoek (NWO), Netherlands;
The Research Council of Norway, Norway;
Commission on Science and Technology for Sustainable Development in the South (COMSATS), Pakistan;
Pontificia Universidad Cat\'{o}lica del Per\'{u}, Peru;
Ministry of Science and Higher Education, National Science Centre and WUT ID-UB, Poland;
Korea Institute of Science and Technology Information and National Research Foundation of Korea (NRF), Republic of Korea;
Ministry of Education and Scientific Research, Institute of Atomic Physics and Ministry of Research and Innovation and Institute of Atomic Physics, Romania;
Joint Institute for Nuclear Research (JINR), Ministry of Education and Science of the Russian Federation, National Research Centre Kurchatov Institute, Russian Science Foundation and Russian Foundation for Basic Research, Russia;
Ministry of Education, Science, Research and Sport of the Slovak Republic, Slovakia;
National Research Foundation of South Africa, South Africa;
Swedish Research Council (VR) and Knut \& Alice Wallenberg Foundation (KAW), Sweden;
European Organization for Nuclear Research, Switzerland;
Suranaree University of Technology (SUT), National Science and Technology Development Agency (NSDTA) and Office of the Higher Education Commission under NRU project of Thailand, Thailand;
Turkish Atomic Energy Agency (TAEK), Turkey;
National Academy of  Sciences of Ukraine, Ukraine;
Science and Technology Facilities Council (STFC), United Kingdom;
National Science Foundation of the United States of America (NSF) and United States Department of Energy, Office of Nuclear Physics (DOE NP), United States of America.
\end{acknowledgement}

\bibliographystyle{utphys}
\bibliography{biblio}
\newpage
\appendix
  
%
%

\section{The ALICE Collaboration}
\label{app:collab}

\begingroup
\small
\begin{flushleft}
S.~Acharya\Irefn{org141}\And 
D.~Adamov\'{a}\Irefn{org95}\And 
A.~Adler\Irefn{org74}\And 
J.~Adolfsson\Irefn{org81}\And 
M.M.~Aggarwal\Irefn{org100}\And 
G.~Aglieri Rinella\Irefn{org34}\And 
M.~Agnello\Irefn{org30}\And 
N.~Agrawal\Irefn{org10}\textsuperscript{,}\Irefn{org54}\And 
Z.~Ahammed\Irefn{org141}\And 
S.~Ahmad\Irefn{org16}
% ...
\And
\renewcommand\labelenumi{\textsuperscript{\theenumi}~}

\section*{Affiliation notes}
\renewcommand\theenumi{\roman{enumi}}
\begin{Authlist}
\item \Adef{org*}Deceased
\item \Adef{orgI}Italian National Agency for New Technologies, Energy and Sustainable Economic Development (ENEA), Bologna, Italy
\item \Adef{orgII}Dipartimento DET del Politecnico di Torino, Turin, Italy
\item \Adef{orgIII}M.V. Lomonosov Moscow State University, D.V. Skobeltsyn Institute of Nuclear, Physics, Moscow, Russia
\item \Adef{orgIV}Department of Applied Physics, Aligarh Muslim University, Aligarh, India
\item \Adef{orgV}Institute of Theoretical Physics, University of Wroclaw, Poland
\end{Authlist}

\section*{Collaboration Institutes}
\renewcommand\theenumi{\arabic{enumi}~}
\begin{Authlist}
\item \Idef{org1}A.I. Alikhanyan National Science Laboratory (Yerevan Physics Institute) Foundation, Yerevan, Armenia
\item \Idef{org2}Bogolyubov Institute for Theoretical Physics, National Academy of Sciences of Ukraine, Kiev, Ukraine
\item \Idef{org3}Bose Institute, Department of Physics  and Centre for Astroparticle Physics and Space Science (CAPSS), Kolkata, India
\item \Idef{org4}Budker Institute for Nuclear Physics, Novosibirsk, Russia
\item \Idef{org5}California Polytechnic State University, San Luis Obispo, California, United States
\item \Idef{org6}Central China Normal University, Wuhan, China
\item \Idef{org7}Centre de Calcul de l'IN2P3, Villeurbanne, Lyon, France
\item \Idef{org8}Centro de Aplicaciones Tecnol\'{o}gicas y Desarrollo Nuclear (CEADEN), Havana, Cuba
\item \Idef{org9}Centro de Investigaci\'{o}n y de Estudios Avanzados (CINVESTAV), Mexico City and M\'{e}rida, Mexico
\item \Idef{org10}Centro Fermi - Museo Storico della Fisica e Centro Studi e Ricerche ``Enrico Fermi', Rome, Italy
\item \Idef{org11}Chicago State University, Chicago, Illinois, United States
\item \Idef{org12}China Institute of Atomic Energy, Beijing, China
\item \Idef{org13}Comenius University Bratislava, Faculty of Mathematics, Physics and Informatics, Bratislava, Slovakia
\item \Idef{org14}COMSATS University Islamabad, Islamabad, Pakistan
\item \Idef{org15}Creighton University, Omaha, Nebraska, United States
\item \Idef{org16}Department of Physics, Aligarh Muslim University, Aligarh, India
\item \Idef{org17}Department of Physics, Pusan National University, Pusan, Republic of Korea
\item \Idef{org18}Department of Physics, Sejong University, Seoul, Republic of Korea
\item \Idef{org19}Department of Physics, University of California, Berkeley, California, United States
\item \Idef{org20}Department of Physics, University of Oslo, Oslo, Norway
\item \Idef{org21}Department of Physics and Technology, University of Bergen, Bergen, Norway
\item \Idef{org22}Dipartimento di Fisica dell'Universit\`{a} 'La Sapienza' and Sezione INFN, Rome, Italy
\item \Idef{org23}Dipartimento di Fisica dell'Universit\`{a} and Sezione INFN, Cagliari, Italy
\item \Idef{org24}Dipartimento di Fisica dell'Universit\`{a} and Sezione INFN, Trieste, Italy
\item \Idef{org25}Dipartimento di Fisica dell'Universit\`{a} and Sezione INFN, Turin, Italy
\item \Idef{org26}Dipartimento di Fisica e Astronomia dell'Universit\`{a} and Sezione INFN, Bologna, Italy
\item \Idef{org27}Dipartimento di Fisica e Astronomia dell'Universit\`{a} and Sezione INFN, Catania, Italy
\item \Idef{org28}Dipartimento di Fisica e Astronomia dell'Universit\`{a} and Sezione INFN, Padova, Italy
\item \Idef{org29}Dipartimento di Fisica `E.R.~Caianiello' dell'Universit\`{a} and Gruppo Collegato INFN, Salerno, Italy
\item \Idef{org30}Dipartimento DISAT del Politecnico and Sezione INFN, Turin, Italy
\item \Idef{org31}Dipartimento di Scienze e Innovazione Tecnologica dell'Universit\`{a} del Piemonte Orientale and INFN Sezione di Torino, Alessandria, Italy
\item \Idef{org32}Dipartimento di Scienze MIFT, Universit\`{a} di Messina, Messina, Italy
\item \Idef{org33}Dipartimento Interateneo di Fisica `M.~Merlin' and Sezione INFN, Bari, Italy
\item \Idef{org34}European Organization for Nuclear Research (CERN), Geneva, Switzerland
\item \Idef{org35}Faculty of Electrical Engineering, Mechanical Engineering and Naval Architecture, University of Split, Split, Croatia
\item \Idef{org36}Faculty of Engineering and Science, Western Norway University of Applied Sciences, Bergen, Norway
\item \Idef{org37}Faculty of Nuclear Sciences and Physical Engineering, Czech Technical University in Prague, Prague, Czech Republic
\item \Idef{org38}Faculty of Science, P.J.~\v{S}af\'{a}rik University, Ko\v{s}ice, Slovakia
\item \Idef{org39}Frankfurt Institute for Advanced Studies, Johann Wolfgang Goethe-Universit\"{a}t Frankfurt, Frankfurt, Germany
\item \Idef{org40}Fudan University, Shanghai, China
\item \Idef{org41}Gangneung-Wonju National University, Gangneung, Republic of Korea
\item \Idef{org42}Gauhati University, Department of Physics, Guwahati, India
\item \Idef{org43}Helmholtz-Institut f\"{u}r Strahlen- und Kernphysik, Rheinische Friedrich-Wilhelms-Universit\"{a}t Bonn, Bonn, Germany
\item \Idef{org44}Helsinki Institute of Physics (HIP), Helsinki, Finland
\item \Idef{org45}High Energy Physics Group,  Universidad Aut\'{o}noma de Puebla, Puebla, Mexico
\item \Idef{org46}Hiroshima University, Hiroshima, Japan
\item \Idef{org47}Hochschule Worms, Zentrum  f\"{u}r Technologietransfer und Telekommunikation (ZTT), Worms, Germany
\item \Idef{org48}Horia Hulubei National Institute of Physics and Nuclear Engineering, Bucharest, Romania
\item \Idef{org49}Indian Institute of Technology Bombay (IIT), Mumbai, India
\item \Idef{org50}Indian Institute of Technology Indore, Indore, India
\item \Idef{org51}Indonesian Institute of Sciences, Jakarta, Indonesia
\item \Idef{org52}INFN, Laboratori Nazionali di Frascati, Frascati, Italy
\item \Idef{org53}INFN, Sezione di Bari, Bari, Italy
\item \Idef{org54}INFN, Sezione di Bologna, Bologna, Italy
\item \Idef{org55}INFN, Sezione di Cagliari, Cagliari, Italy
\item \Idef{org56}INFN, Sezione di Catania, Catania, Italy
\item \Idef{org57}INFN, Sezione di Padova, Padova, Italy
\item \Idef{org58}INFN, Sezione di Roma, Rome, Italy
\item \Idef{org59}INFN, Sezione di Torino, Turin, Italy
\item \Idef{org60}INFN, Sezione di Trieste, Trieste, Italy
\item \Idef{org61}Inha University, Incheon, Republic of Korea
\item \Idef{org62}Institute for Nuclear Research, Academy of Sciences, Moscow, Russia
\item \Idef{org63}Institute for Subatomic Physics, Utrecht University/Nikhef, Utrecht, Netherlands
\item \Idef{org64}Institute of Experimental Physics, Slovak Academy of Sciences, Ko\v{s}ice, Slovakia
\item \Idef{org65}Institute of Physics, Homi Bhabha National Institute, Bhubaneswar, India
\item \Idef{org66}Institute of Physics of the Czech Academy of Sciences, Prague, Czech Republic
\item \Idef{org67}Institute of Space Science (ISS), Bucharest, Romania
\item \Idef{org68}Institut f\"{u}r Kernphysik, Johann Wolfgang Goethe-Universit\"{a}t Frankfurt, Frankfurt, Germany
\item \Idef{org69}Instituto de Ciencias Nucleares, Universidad Nacional Aut\'{o}noma de M\'{e}xico, Mexico City, Mexico
\item \Idef{org70}Instituto de F\'{i}sica, Universidade Federal do Rio Grande do Sul (UFRGS), Porto Alegre, Brazil
\item \Idef{org71}Instituto de F\'{\i}sica, Universidad Nacional Aut\'{o}noma de M\'{e}xico, Mexico City, Mexico
\item \Idef{org72}iThemba LABS, National Research Foundation, Somerset West, South Africa
\item \Idef{org73}Jeonbuk National University, Jeonju, Republic of Korea
\item \Idef{org74}Johann-Wolfgang-Goethe Universit\"{a}t Frankfurt Institut f\"{u}r Informatik, Fachbereich Informatik und Mathematik, Frankfurt, Germany
\item \Idef{org75}Joint Institute for Nuclear Research (JINR), Dubna, Russia
\item \Idef{org76}Korea Institute of Science and Technology Information, Daejeon, Republic of Korea
\item \Idef{org77}KTO Karatay University, Konya, Turkey
\item \Idef{org78}Laboratoire de Physique des 2 Infinis, Ir\`{e}ne Joliot-Curie, Orsay, France
\item \Idef{org79}Laboratoire de Physique Subatomique et de Cosmologie, Universit\'{e} Grenoble-Alpes, CNRS-IN2P3, Grenoble, France
\item \Idef{org80}Lawrence Berkeley National Laboratory, Berkeley, California, United States
\item \Idef{org81}Lund University Department of Physics, Division of Particle Physics, Lund, Sweden
\item \Idef{org82}Nagasaki Institute of Applied Science, Nagasaki, Japan
\item \Idef{org83}Nara Women{'}s University (NWU), Nara, Japan
\item \Idef{org84}National and Kapodistrian University of Athens, School of Science, Department of Physics , Athens, Greece
\item \Idef{org85}National Centre for Nuclear Research, Warsaw, Poland
\item \Idef{org86}National Institute of Science Education and Research, Homi Bhabha National Institute, Jatni, India
\item \Idef{org87}National Nuclear Research Center, Baku, Azerbaijan
\item \Idef{org88}National Research Centre Kurchatov Institute, Moscow, Russia
\item \Idef{org89}Niels Bohr Institute, University of Copenhagen, Copenhagen, Denmark
\item \Idef{org90}Nikhef, National institute for subatomic physics, Amsterdam, Netherlands
\item \Idef{org91}NRC Kurchatov Institute IHEP, Protvino, Russia
\item \Idef{org92}NRC \guillemotleft Kurchatov\guillemotright~Institute - ITEP, Moscow, Russia
\item \Idef{org93}NRNU Moscow Engineering Physics Institute, Moscow, Russia
\item \Idef{org94}Nuclear Physics Group, STFC Daresbury Laboratory, Daresbury, United Kingdom
\item \Idef{org95}Nuclear Physics Institute of the Czech Academy of Sciences, \v{R}e\v{z} u Prahy, Czech Republic
\item \Idef{org96}Oak Ridge National Laboratory, Oak Ridge, Tennessee, United States
\item \Idef{org97}Ohio State University, Columbus, Ohio, United States
\item \Idef{org98}Petersburg Nuclear Physics Institute, Gatchina, Russia
\item \Idef{org99}Physics department, Faculty of science, University of Zagreb, Zagreb, Croatia
\item \Idef{org100}Physics Department, Panjab University, Chandigarh, India
\item \Idef{org101}Physics Department, University of Jammu, Jammu, India
\item \Idef{org102}Physics Department, University of Rajasthan, Jaipur, India
\item \Idef{org103}Physikalisches Institut, Eberhard-Karls-Universit\"{a}t T\"{u}bingen, T\"{u}bingen, Germany
\item \Idef{org104}Physikalisches Institut, Ruprecht-Karls-Universit\"{a}t Heidelberg, Heidelberg, Germany
\item \Idef{org105}Physik Department, Technische Universit\"{a}t M\"{u}nchen, Munich, Germany
\item \Idef{org106}Politecnico di Bari, Bari, Italy
\item \Idef{org107}Research Division and ExtreMe Matter Institute EMMI, GSI Helmholtzzentrum f\"ur Schwerionenforschung GmbH, Darmstadt, Germany
\item \Idef{org108}Rudjer Bo\v{s}kovi\'{c} Institute, Zagreb, Croatia
\item \Idef{org109}Russian Federal Nuclear Center (VNIIEF), Sarov, Russia
\item \Idef{org110}Saha Institute of Nuclear Physics, Homi Bhabha National Institute, Kolkata, India
\item \Idef{org111}School of Physics and Astronomy, University of Birmingham, Birmingham, United Kingdom
\item \Idef{org112}Secci\'{o}n F\'{\i}sica, Departamento de Ciencias, Pontificia Universidad Cat\'{o}lica del Per\'{u}, Lima, Peru
\item \Idef{org113}St. Petersburg State University, St. Petersburg, Russia
\item \Idef{org114}Stefan Meyer Institut f\"{u}r Subatomare Physik (SMI), Vienna, Austria
\item \Idef{org115}SUBATECH, IMT Atlantique, Universit\'{e} de Nantes, CNRS-IN2P3, Nantes, France
\item \Idef{org116}Suranaree University of Technology, Nakhon Ratchasima, Thailand
\item \Idef{org117}Technical University of Ko\v{s}ice, Ko\v{s}ice, Slovakia
\item \Idef{org118}The Henryk Niewodniczanski Institute of Nuclear Physics, Polish Academy of Sciences, Cracow, Poland
\item \Idef{org119}The University of Texas at Austin, Austin, Texas, United States
\item \Idef{org120}Universidad Aut\'{o}noma de Sinaloa, Culiac\'{a}n, Mexico
\item \Idef{org121}Universidade de S\~{a}o Paulo (USP), S\~{a}o Paulo, Brazil
\item \Idef{org122}Universidade Estadual de Campinas (UNICAMP), Campinas, Brazil
\item \Idef{org123}Universidade Federal do ABC, Santo Andre, Brazil
\item \Idef{org124}University of Cape Town, Cape Town, South Africa
\item \Idef{org125}University of Houston, Houston, Texas, United States
\item \Idef{org126}University of Jyv\"{a}skyl\"{a}, Jyv\"{a}skyl\"{a}, Finland
\item \Idef{org127}University of Liverpool, Liverpool, United Kingdom
\item \Idef{org128}University of Science and Technology of China, Hefei, China
\item \Idef{org129}University of South-Eastern Norway, Tonsberg, Norway
\item \Idef{org130}University of Tennessee, Knoxville, Tennessee, United States
\item \Idef{org131}University of the Witwatersrand, Johannesburg, South Africa
\item \Idef{org132}University of Tokyo, Tokyo, Japan
\item \Idef{org133}University of Tsukuba, Tsukuba, Japan
\item \Idef{org134}Universit\'{e} Clermont Auvergne, CNRS/IN2P3, LPC, Clermont-Ferrand, France
\item \Idef{org135}Universit\'{e} de Lyon, Universit\'{e} Lyon 1, CNRS/IN2P3, IPN-Lyon, Villeurbanne, Lyon, France
\item \Idef{org136}Universit\'{e} de Strasbourg, CNRS, IPHC UMR 7178, F-67000 Strasbourg, France, Strasbourg, France
\item \Idef{org137}Universit\'{e} Paris-Saclay Centre d'Etudes de Saclay (CEA), IRFU, D\'{e}partment de Physique Nucl\'{e}aire (DPhN), Saclay, France
\item \Idef{org138}Universit\`{a} degli Studi di Foggia, Foggia, Italy
\item \Idef{org139}Universit\`{a} degli Studi di Pavia, Pavia, Italy
\item \Idef{org140}Universit\`{a} di Brescia, Brescia, Italy
\item \Idef{org141}Variable Energy Cyclotron Centre, Homi Bhabha National Institute, Kolkata, India
\item \Idef{org142}Warsaw University of Technology, Warsaw, Poland
\item \Idef{org143}Wayne State University, Detroit, Michigan, United States
\item \Idef{org144}Westf\"{a}lische Wilhelms-Universit\"{a}t M\"{u}nster, Institut f\"{u}r Kernphysik, M\"{u}nster, Germany
\item \Idef{org145}Wigner Research Centre for Physics, Budapest, Hungary
\item \Idef{org146}Yale University, New Haven, Connecticut, United States
\item \Idef{org147}Yonsei University, Seoul, Republic of Korea
\end{Authlist}
\endgroup
  
\end{document}